# Time Cycles in Indian Cosmology

Roopa H. Narayan

## 1 Introduction

It is well known that astronomy in ancient cultures began with the recognition of regularity in astronomical phenomena and the determination of various time cycles associated with these phenomena. Santillana and Dechend [1] have shown that ancient myth expresses knowledge of the changing frame of earth's axis. In specific mythologies, there may be additional structure that reveals further information about the knowledge of astronomy [2]. This subject is now a part of the larger discipline of archaeoastronomy.

For analysis from such an archaeoastronomy perspective, Indian texts provide especially rich material due to the antiquity of these texts and the long period over which newer such texts were written. In this article, we review some key time cycles in ancient Indian astronomy, especially those that have emerged from researches in the past couple of decades [3-7].

## 2 Altar astronomy

Kak, in papers published in the *Quarterly Journal of the Royal Astronomical Society* in the 90s, described the astronomy that underlies the altar ritual of the Vedic times most explicitly [3], [4], [12], [13]. The Rig Veda, which is India's earliest extant text and various dated to the $3^{rd}$ or the $2^{nd}$ millennium BC, discusses rituals and sacrifices using altars which have an astronomical basis. The underlying principle used is a tripartite division of the cosmos as well as the individual and equivalence between the two is made to present data that related to both.

In an ancient altar ritual the symbolic year of 360 days which represents the universe is divided in to three parts representing three divisions:

| Sky | Space | Earth | |
|-----|-------|-------|---|
| 261 | 78 | 21 | 268+78+21= 360days |

The basis of this appears to be related to the identification of Vishnu with Mercury and the three steps of Vishnu in mythology refer to the three revolutions in 261 days =261/3 = 87 days for each cycle of Mercury around Sun [4]. This division is also central to the astronomical ritual that is described in the Rig Veda.

Furthermore, there is a structure to the very organization of the Rig Veda [5]. In particular, if the 10 books of the Rig Veda are viewed as a 5-layered altar, with two books to each layer in a specific arrangement, then we find various regularities in the number of hymns that are in the various books. This organization appears to be based on ideas of equivalence relating the year and its various subdivisions to the text of the Veda.

## 2.1 The Agnicayana altar

One of the well known rituals in the Vedic times is the *Agnicayana* ritual. The altar of this ritual and the mathematics and astronomy coded in it gives an idea of the Vedic rituals. The altar construction consists of five layers built in bricks in the shape of a falcon. This altar represents all the three types of years: *nakshatra* (constellation), solar and lunar. The Vedic people used three different kinds of calculations for years depending on where such a year had to be used. The solar year is the time taken by the earth to go around sun, lunar is defined by moon, and constellation is related to stars [5].

The *Agnicayana* altar with five layers and their representation:

I layer – represents earth

II layer – represents merging of earth and space

III layer – represents Space

IV layer – represents merging of space and sky

V layer – represents sky

Yajnavalkya describes fire altars as representatives of both knowledge and ritual. The bricks used in the construction of these altars are specially designed and just like the layers of the altar structure having a hidden meaning even the number of bricks used are associated with astronomical knowledge. The total number of bricks used in this altar is 10,800 which is equal to the number of *muhurtas* in a year and each day has 30 *muhurtas*: 10,800 = 360x30.

Layer I –    98 bricks

Layer II -   41 bricks

    |——41+71=112, 1/3 number of days in a *nakshatra* year (=28x12=336)

Layer III -  71 bricks

    |——71+47=118, integer nearest to 1/3 of lunar year

Layer IV -   47 bricks

    |——47+138=186, tithis in the half-year

Layer V  -  138 bricks

It is quite clear from these numbers that they are related to the lengths of the three different measures of the year or their subdivisions.

## 3 Orion

Tilak [2] argued convincingly that the *Upanayana* (initiation of boys) ritual had elements that described the knowledge of the shifting of the seasons away from Orion (Prajapati in Sanskrit).

In Sayana's commentary on Shatapatha Brahmana the reference to Prajapati is as:

इषुणा तस्य शिरश्चिच्छेद ...... इषुः शिरश्चेत्युभयमंतरिक्षमुत्प्लुत्य नक्षत्रात्मनावस्थितं दृश्यते। S.Br .ii. 1. 2. 8.

In this Sayana mentions that Rudra – the form created by gods "cut off Prajapati's head by the arrow", "the arrow and the head both jumped up to the heavens and are there stationed".

In the *Aitareya Brahmana* and the Rig the same story or rather the same event is referred to with changes in the mythical characters, at times it is Indra who cuts off the head of Vritra or in a different source it is the head of *Mrighashiras*/antelope cut off by Indra [2].

Tilak argues for the fact that the hymn in Rig Veda which talks about the dogs which possess four eyes and guard Yama's region beginning the new year actually refers to Canis near the milky way with the four stars as the four eyes. Further in the Mahimna stotra Rudra takes a place in the sky and the beauty of Ganga on his head is enhanced by a number of stars. This Rudra pierces Prajapati with an arrow and the whole story is said to be illustrated in the sky. There is an elaborate argument with added support from Greek and Parsi texts besides the Puranas.

Interestingly the story does not end here. Prajapati who is also called *Yajna* or the sacrifice which itself has astronomical significance as shown elsewhere has a girdle or piece of cloth around the waist.

Because it is Orion or Prajapati or the *Yajna* who has the piece of cloth, as all words are deductive in Sanskrit, this piece of cloth is called '*Yajnopavita*' which means the *upavitha* of *Yajna* or the cloth of *Yajna*.

This is supported by a verse found in Brahmopanishad but not found in Samhitas which may be due to the forgetting of the significance over time.

यज्ञोपवीतं परमं पवित्रं प्रजापतेर्यत्सहजं पुरस्तात् ।
This means that *Yajnopavita* is high and sacred and was born with Prajapati of the old.

Understanding this in the context of *Upanayan*a ceremony where the one who is subject to it has to wear the *mekhala* grass – a kind of grass around the waist with three knots representing the three stars of Orion as it is mentioned that the knots are dear to soma who is the presiding deity of Orion is important. Also this person has to carry a stick or staff accompanied by deer skin and the thread of *yajnapovita*. These three *mekhala, ajina* and *danda*, i.e., the staff, deer skin and three knots seem to be used to imitate the costume of Prajapati [2].

It is also worth wondering whether the necklace with three significant knots tied in the neck of the bride in the marriage ceremony is another form of the same as above since even the *yajnopavita* was known to be worn as a necklace.

## 4 Temple cosmology

As in the sacred architecture of other cultures, the very structure of the Hindu temples is designed to reflect cosmological knowledge [8-11], [14-18].

Specifically, 108 is a very sacred number in the Indian tradition. The number of beads in a japamala or the rosary bead string used for meditation is 108. the number of dance poses described in Natya Shastra an authoritative text on Dance is 108. The total number of syllables in Rig Veda is 432,000 which is divisible of 108. The distance between sun and earth and moon and earth in terms of their respective diameter is 108 too [19].

The temples seem to be more of observatory and less of a religious place when their design is examined in detail. The rising sun on equinox and solstice days was aligned on the western entrance of the temple. Many sighting lines for the sunrise and moonrise are identifiable from the temple.

For illustration, we consider the Vishnu temple of Angkor Wat in Cambodia. In the central tower of this temple [20], the top most elevation has external axial dimensions of 189.00 cubit east-west and 176.37 cubit north-south. The total sum of these numbers is 365.37 which is very close to the count of days in a solar year and the unequal division of this number through dimensions is the unequal divisions of a year due to the asymmetric motion of sun as mentioned in the Shatapatha Brahmana [11].

The different measurements in the temple correspond to 365 days of a year, 360 tithis of a year, 27 nakshatras / constellations, etc.

The west-east axis in the temple represents the Yugas. The following are the four measured between two points in the latter three cases and just a measurement in the first case:

| | |
|---|---|
| The width of the moat | = 439.78cubits |
| First step of western entrance gateway to balustrade wall | = 867.03cubits |
| First step of western entrance gateway to first step of central tower | = 1,296.07cubits |
| First step of bridge to geographic center of temple | = 1,734.41cubits |

These four numbers correspond to the years in the four *yugas* in years:

   *Kali Yuga*      = 432,000
   *Dvapara Yuga* = 864,000
   *Treta Yuga*    = 1,296,000
   *Krita Yga*     = 1,728,000

There is evidence that the cycles of the *yugas* emerged out of the desire to harmonize the varying periods of the sun, the moon, and the planets. A *kalpa* was taken to constitute 1000 *Chaturyugas* = 4 *yugapadas* = 4,320,000 sidereal years.

1 *Kalpa* = 1000 *Chaturyugas*
      = 4,320,000,000 sidereal years

A day of Brahma is = A night of Brahma = 1 *Kalpa* = 4320000000 sidereal years.
At the end of one Brahma year which has 360 Brahma days and 360 Brahma nights the whole cosmos is re-created. This is the complete state of chaos and there is a rest period of One Brahma year after the cosmic dissolution before a new cosmic creation begins.

One day of Brahma     4320000000 sidereal years
One night of Brahma   4320000000 sidereal years
                        ---------------
                        8,640,000,000 sidereal years

Days in a Brahma year   - 360days

One year of Brahma in sidereal years is
$360 \times 8640000000 = 3.1104 \times 10^{12}$ sidereal years

Therefore 100 Brahma years is
$100 \times 3.1104 \times 10^{12}$ sidereal years $= 3.1104 \times 10^{14}$ sidereal years

Therefore
Brahma's Life = $3.1104 \times 10^{14}$ sidereal years.

This whole cycle of creation and decay is supposed to follow a cosmic law called *Rta* – the law of universe.

Indian cosmology is based on the principles of recursion in terms of the cycles of creation and destruction, pattern repetition at atomic level and higher scales. This further has the implication that the same elements were sought in the outer and the inner cosmos. Furthermore, this explains the importance that was placed on the use of similar ontology for the two.

# 5 The inner cosmos

The cyclic aspects of the inner cosmos are described in Indian texts. The most puzzling aspect, to the materialist, is the observer in the inner cosmos. This observer in Indian context is understood through two main qualities that do not belong to any other life forms which are the mind and the 'awareness/consciousness'. Kanada describes is thus (for background, see [21]):

कारणाज्ञानात्॥३।१।४॥
Because the causes or constituents are devoid of cognition or consciousness.3.1.4

Kanada in his chapter on consciousness (we use this term only for convenience and lack of a better term) first establishes that the object, the perception of the object, and the senses of perception are different. In the sutra stated above he states that the sense organs merely act as intermediate agents between an object and the consciousness of the observer. E.g. hand by itself cannot be aware of a computer for in the absence of hand too the computer is necessarily perceived. Therefore consciousness must be ability of an observer to connect to reality in a certain way.

How is this observer any different from animals? We will examine this as we go on.

अज्ञानाच्च॥३।१।६॥
And because it is not known (that any minute degree of consciousness exists in the material objects like chair, etc.) 3.1.6

Kanada himself wonders if the basic entity of matter form *anu*/atom possess this kind of consciousness and through this sutra states that such possibility does not exist even in minute forms since material objects like chair, etc do not exhibit any such abilities. This he says is known.

In the 11 sutras that follow he argues about what is the distinguishing characteristic of the consciousness and explains the different accepted traditional methods of either proving or disproving the presence of an object.

आत्मेन्द्रियार्थसन्निकर्षाद्यन्निष्पद्यते तदन्यत्॥३।१।१८॥
That which is produced from the contact of the consciousness, the sense organs, and the object is other (than a false Mark). 3.1.18

Kanada after discussing how is the presence or absence of something which is not recognizable in any general sense by the effects, causes or sense organs states this just before concluding the presence of consciousness in an observer. In this sutra he is describing the knowledge or a certain kind of awareness or connectedness in a relational sense that is produced with in an observer after perceiving an object.

This kind of connectedness is assumed in relational theories of physics [18], [19]. In this situation to separate the observer from the observed would be breaking that unity.

आत्मेन्द्रियार्थसन्निकर्षे ज्ञानस्य भावोऽभावश्च मनसो लिङ्गम्॥३।२।१॥

The appearance and non-appearance of knowledge, on contact of the consciousness with the senses and their objects, are the characteristics (of the existence) of the mind. 3.2.1

Kanada is associating the mind with knowledge gained which probably is what we refer to as intelligence as associated with the mind today. He separates the mind and the consciousness of the observer may be to state that it is the mind which exists in animals too, though evolved to a different degree.

It is the consciousness which makes an observer feel connected to the universe in a greater way which distinguishes her from all other creatures of the universe. It is this connectedness that is taken to be at the basis of the use of the same models for the inner and the outer cosmos.

## Conclusion

Time cycles in Indian cosmology originally began as the attempt to harmonize the various periods in the motion of the heavenly bodies. The philosophical basis underlying it, beyond the obvious seasonal patterns of the year, was that just as there are cycles in the inner world (the body of the individual), there are corresponding – and longer – cycles in the outer world.

The earliest Indian astronomy (as is found underlying the Vedic altar ritual) was principally concerned with the harmonization of the movements of the sun and the moon. In the further elaboration of this approach, an attempt was made to account for the cycles associated with the various planets. This made it essential to go to much longer periods, into thousands of years rather than the much smaller periods earlier on.

It appears that the knowledge of the precession cycle was the impetus for considering even longer periods that led ultimately, on grounds of symmetry, to the notion of the kalpa of 4,320,000,000 years and Brahma's day of $3.1104 \times 10^{12}$ years.

## References

1. G. de Santillana and H. von Dechend, Hamlet's Mill. Gambit, Boston, 1969.
2. Bal Gangadhar Tilak, *Orion or Researches in to the Antiquity of the Vedas.* Ashtekar & Co., Poona. 1916
3. S. Kak, "The astronomy of the age of geometric altars." Quarterly Journal of the Royal Astronomical Society, vol. 36, pp. 385-396, 1995.
4. S. Kak, "Knowledge of planets in the third millennium BC." Quarterly Journal of the Royal Astronomical Society, vol. 37, pp. 709-715, 1996.

5. S. Kak, *The Astronomical Code of The Rig Veda*. Munshiram Manoharlal, New Delhi, 2000.
6. K.D. Abhyankar, "Uttarayana." In Scientific Heritage of India, B.V.Subbarayappa and S.R.N. Murthy (eds.). The Mythic Society, Bangalore, 1988.
7. B.N. Narahari Achar, "A case for revising the date of Vedanga Jyotisa," Indian Journal of History of Science, vol. 35, pp. 173-183, 2000.
8. S. Kak, Indian Physics: Outline of Early History. 2003; arXiv:physics/0310001
9. S. Kak, Concepts of Space, Time and Consciousness in Ancient India. 1999; arXiv:physics/9903010
10. S. Kak, Greek and Indian Cosmology: Review of Early History. 2006; arXiv: physics/0303001
11. S. Kak, Birth and Early Development of Indian Astronomy, 2001; arXiv: physics/0101063
12. S. Kak, "Astronomy of the Vedic altars." Vistas in Astronomy, vol. 36, pp. 117-140, 1993.
13. S. Kak, "Astronomy in the Vedic altars and the Rgveda." Mankind Quarterly, vol. 33, pp. 43-55 , 1992.
14. Swami Venkatesananda, *The Concise Yoga-Vasistha*. State University of New York Press, albany,1984
15. Dwight W Johnson, *Exegesis of Hindu Cosmological Time Cycles*. 2003.
16. S. Kak, Early theories on the distance to the Sun, 1998; arXiv: physics/9804021
17. S. Kak, *The Nature of Physical Reality*. Peter Lang, New York, 1986.
18. S. Kak, "On the science of consciousness in ancient India." Indian Journal of History of Science, vol 32, 1997, 105-120.
19. S. Kak, *The Architecture of Knowledge*. Center for Studies in Civilizations/ Motilal Banarsidass, New Delhi, 2004.
20. E. Mannikka, *Angkor Wat: Time, Space and Kingship*. Univ of Hawaii Press, Honolulu, 1996.
21. R. H. Narayan, Indian cosmological ideas, 2006. arXiv: arXiv:0705.1192v1